\documentclass{PoS}

\usepackage{color}
\usepackage{lineno}
\usepackage{deluxetable}
\usepackage{xcolor}%

\title{Measurement of the Diffuse Astrophysical Muon-Neutrino Spectrum with Ten Years of IceCube Data}

\ShortTitle{IceCube: Diffuse Astrophysical Muon-Neutrino Spectrum}

\author{
The IceCube Collaboration\footnote{For collaboration list, see PoS(ICRC2019) 1177.}\\
{\itshape \href{http://icecube.wisc.edu/collaboration/authors/icrc19_icecube}{http://icecube.wisc.edu/collaboration/authors/icrc19\_icecube}}\\
E-mail: \email{stettner@physik.rwth-aachen.de}
}

\abstract{The IceCube Neutrino Observatory measured a flux of high-energy astrophysical neutrinos in several detection channels. The energy spectrum is fitted as unbroken power-law, but different best-fit parameters were obtained in the various analyses covering different energy ranges between few TeV to 10 PeV. Here, we present an update to the analysis of through-going muon-neutrinos from the Northern Hemisphere. It was extended to almost ten years of data and an improved treatment of systematic uncertainties on the atmospheric fluxes was implemented. The updated best-fit parameters for the astrophysical flux assuming a power-law energy spectrum are $\Phi_{astro}=1.44$ and $\gamma_{astro}=2.28$. We will present the results of the spectral fit and discuss how the measured flux compares to other IceCube results. \\

\vspace{4mm}
{\bfseries Corresponding authors:}
\speaker{J. Stettner}$^{1}$\\
{$^{1}$ \itshape III. Physikalisches Institut, RWTH Aachen University, D-52056 Aachen, Germany} 
}

\FullConference{36th International Cosmic Ray Conference -ICRC2019-\\
		July 24th - August 1st, 2019\\
		Madison, WI, U.S.A.}

\begin{document}

\section{Introduction}
In 2013 the IceCube Neutrino Observatory announced the discovery of a high-energy astrophysical neutrino flux~\cite{Aartsen:2013jdh}. This was accomplished by using an outer veto to reduce the atmospheric background and enable the study of highest-energy starting events.
Shortly after, the discovery was confirmed in the complementary channel of through-going and starting muons coming from the Northern Hemisphere where the atmospheric muon background is suppressed~\cite{Aartsen:2015rwa,Aartsen:2016xlq}. Since then, increasingly larger samples of through-going muon-neutrinos have been used to measure the high-energy astrophysical flux. 

Here, we report on the latest analysis of northern sky muon-neutrinos encompassing almost ten years of data-taking. The event selection is unchanged compared to previous iterations of the analysis~\cite{Aartsen:2016xlq}: It restricts itself to the observation of the Northern Celestial Hemisphere and identifies high-quality tracks based on a boosted decision tree. Because a median angular resolution of better than $\sim 1^{\circ} (E_\nu \geq 1\,\mathrm{TeV})$ is achieved for these events, (mis-reconstructed) atmospheric muons can be rejected efficiently resulting in a 99.7\% purity of the sample. 

\section{The IceCube Neutrino Observatory and the Data Sample}

The IceCube Neutrino Observatory is a cubic-kilometer Cherenkov detector embedded in the Antarctic ice 
at the South Pole~\cite{Aartsen:2016nxy}. It detects neutrinos by observing Cherenkov radiation emitted by charged secondary particles that are created in neutrino interactions. A total of 5160 optical sensors instrument 86 vertical cables, often called strings, located beneath the surface which are arranged in an active volume of about one cubic kilometer.

%at depths between 1450m and 2450m 
\begin{figure}[t]
	\centering
	\includegraphics[width=0.75\columnwidth]{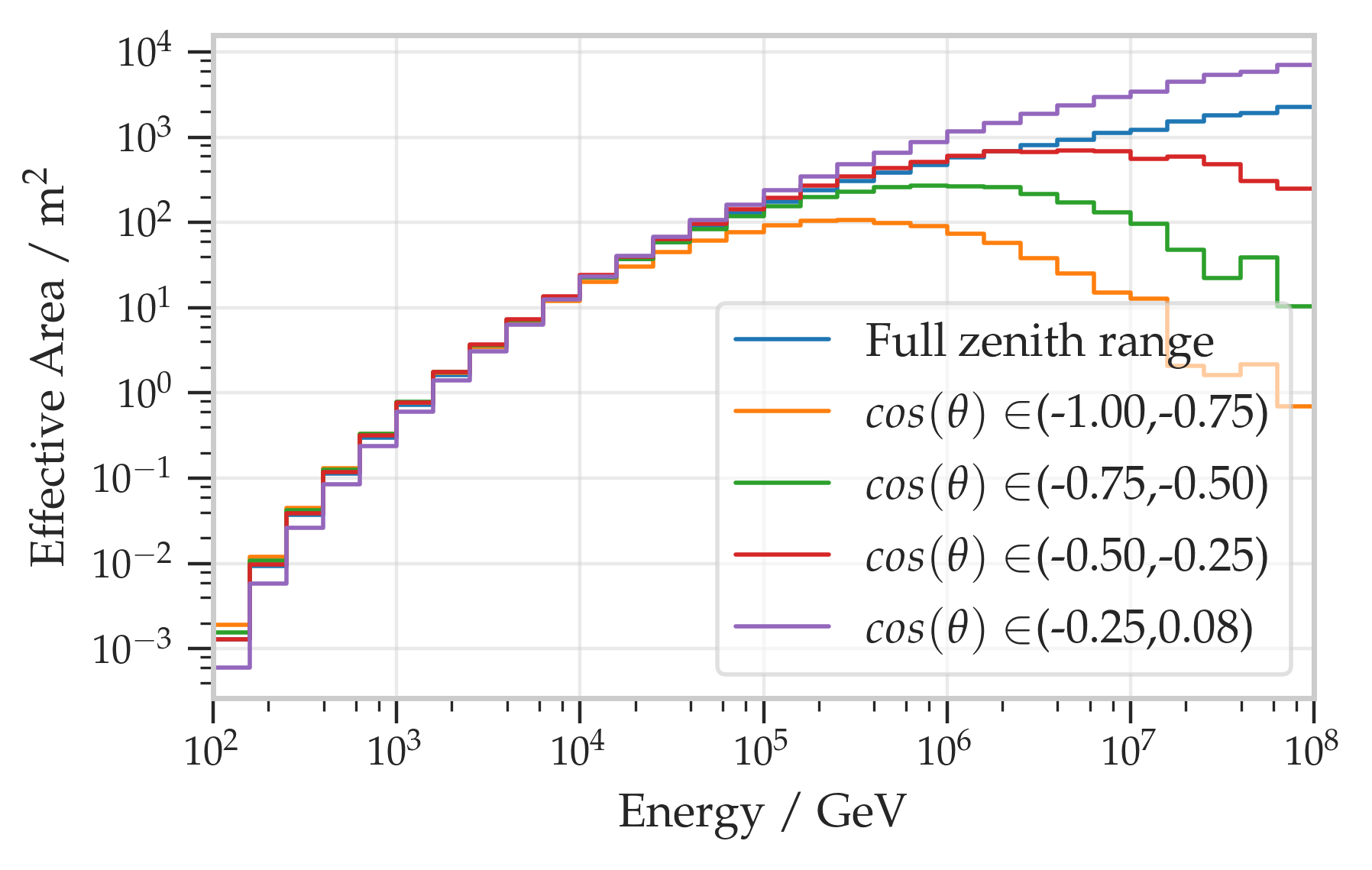}
   \caption{Total muon-neutrino ($\nu_\mu + \bar{\nu}_{\mu}$) effective area of the event selection. Different zenith ranges are shown as a function of the incident neutrino energy. \label{fig:eff-area}}
\end{figure}

IceCube was completed in December 2010, but data-taking was on-going in previous years with partial detector configurations. The analysis presented here uses data from May 2009 to December 2018, taken with the 59-string configuration --short-handed IC59-- in the first year, the 79-string configuration --labeled IC79-- in the second year, and the 86-string configuration of the completed detector afterwards labeled as IC86. 

%It was realized that the charge-scale as derived from single photo-electrons was shifted by about 
An improved description of the observed charge distribution for single photo-electrons has resulted in a $\sim 4\%$ shift of the charge-scale for historical data. To correct for this, a large re-calibration campaign, called Pass-2, was conducted and the experimental data from the years 2010 to 2016 have been re-analyzed. Within this effort, the same filtering, event selection and reconstructions have been applied consistently to all events, except for IC59. As a consequence, the energy and directional reconstruction of some events changed compared to the previous publication. For the data taken after May 2016, the correct calibration has been applied as default. 
Figure \ref{fig:eff-area} shows the updated effective area of the event selection as a function of neutrino energy. The total number of events can be obtained by integrating the product of effective area, livetime and average of the neutrino and anti-neutrino flux $(\phi_{\nu_\mu} + \phi_{\bar{\nu}_\mu}) / 2$ over neutrino energy and solid angle.

The full dataset now consists of about 650,000 neutrino events, including approximately 150,000 new events compared to the last iteration of this analysis~\cite{ICRC17NuMu}. Notably, one new event with a reconstructed muon energy of $E_{\mathrm{\mu, reco}} = 1.2\,\mathrm{PeV}$ was observed on November 6th 2017. This event was also reported as a real-time alert~\cite{GNCalert_EHE20171106}.

\begin{table}
\centering
{\small
\begin{tabular}{lcccc}
\hline
{Data-taking Season} & {Zenith-range / deg} & {Livetime / sec} & {Number of events} & {Pass-2}\\
\hline
         IC59               & $90 - 180$ &  $30079123$ & 21411 & No \\
         IC79 -- IC86-2018  & $85 - 180$ & $259620998$ & 630341 & Yes \\
        \hline
\end{tabular}\\
}
\caption{Summary of the data used in this analysis. Each row corresponds to a detector configuration. 
%Analysis Zenith ranges, livetimes and number of observed events are shown in each column. 
See text for a description of the Pass-2 re-calibration campaign that was conducted for all data after May 2010. \label{tab:seasons}}
\end{table}

\section{Analysis Method}
\subsection{Neutrino Flux Models}
\label{sec:neutrinofluxes}
We consider three contributions to the total neutrino flux: conventional atmospheric neutrinos from the decay of pions and kaons in cosmic-ray induced air showers, prompt atmospheric neutrinos from the decay of charmed hadrons in the same showers, and an isotropic flux of astrophysical neutrinos. 

The conventional and prompt atmospheric fluxes are computed using the Matrix-Cascade-Equation solver package MCEq~\cite{Fedynitch:2015zma}. Calculations are performed with the MSIS-00~\cite{MSIS00_2002JGRA} atmospheric model and the SIBYLL2.3c model of hadronic interactions~\cite{Fedynitch:2018cbl_SIBYLL23c}. The Gaisser-Hillas model (H4a)~\cite{Gaisser:2012_H3a_n_H4a} is used as baseline for the primary cosmic-ray flux, but variations are implemented as a nuisance parameter in the fit to account for systematic uncertainties.

As baseline spectral model for the astrophysical flux, a single power-law is assumed:
\begin{equation}
    \frac{d\phi}{dE} = \Phi_{astro} \times \left ( \frac{E_\nu}{100\,\mathrm{TeV}} \right )^{-\gamma_{astro}},
\end{equation}
where $\Phi_{astro}$ is the normalization of the astrophysical component and $\gamma_{astro}$ the spectral slope.

\subsection{Statistical Methodology}
The analysis is based on comparing the expected astrophysical signal as well as atmospheric background expectation as obtained from Monte Carlo simulations to the experimental data. The data and Monte Carlo are binned in two observables: estimated muon energy and cosine of the reconstructed muon zenith angle. Observed counts are compared to the expectation in each bin of the resulting histogram using the Poisson likelihood $\mathcal{L}_{i}$.
%The expected counts in each bin of the resulting histograms is modeled by a Poisson distribution $\mathcal{L}_{i}$.
The bin-wise expectation $\mu_i$ is defined as:
\begin{equation}
\mu_i(\mathbf{\theta}, \mathbb{\xi}) =  \mu_i^{conv.}(\xi)  +\mu_i^{prompt}(\Phi_{prompt}, \xi) + \mu_i^{astro}(\Phi_{astro}, \gamma_{astro}, \xi_{det.}),
\end{equation}
where $\theta$ denotes the signal flux parameters $\Phi_{astro}, \gamma_{astro}$, %, \Phi_{prompt}$, 
and $\xi = \{\xi_{bkg.}; \xi_{det.}\}$ denotes the nuisance parameters. The latter incorporate systematic uncertainties into the likelihood function. The total likelihood is obtained as $\mathcal{L} = \prod_{i = 1}^{n_{\mathrm{bins}}} ~\mathcal{L}_i$ where ${n_{\mathrm{bins}}}$ is the number of bins. Best-fit signal and nuisance parameters are obtained by maximizing the total likelihood function. Additionally, parameter confidence regions are calculated from the profile likelihood over all other parameters applying Wilks' theorem \cite{bibWilks}.
%The applicability of Wilks' theorem has been verified using ensemble tests.

\subsection{Background and Detector Uncertainties}
\begin{figure}[htbp]
        \centering
		\includegraphics[width=0.94\textwidth]{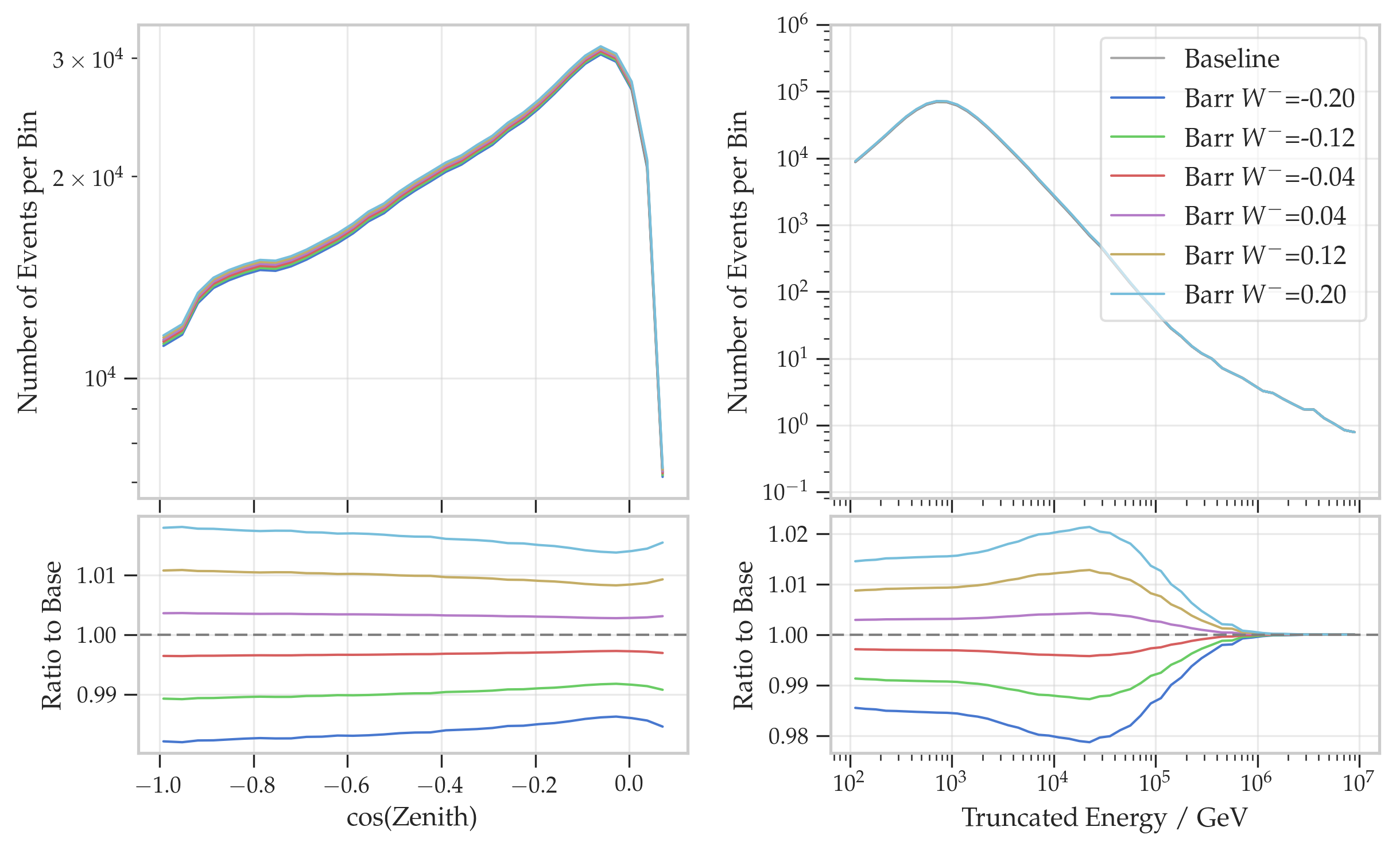}
\caption{Impact of varied nuisance parameters on the total flux expectation: As an example, variations of the $W^{-}$-parameter as defined in Barr et al.~\cite{PhysRevD.74.094009} are shown, all other fit-parameters are set to their baseline value.} 
    \label{fig:barrsystematics}
\end{figure}

We consider systematic uncertainties on the atmospheric background predictions 
($\xi_{bkg.}$) and detector effects ($\xi_{det.}$), both are included as continuous parameters in the analysis. A detailed description of the detector uncertainties and their technical implementation can be found in \cite{Aartsen:2016xlq}. The treatment of atmospheric flux uncertainties has been improved by including the variation of primary cosmic-ray models and parameterizing the hadronic interaction uncertainties by following the approach in Barr et al.~\cite{PhysRevD.74.094009}. In this approach different parts of the phase space for hadron production in proton-air collisions are separated and the production rate of these hadrons is scaled independently: The eight parameters $H^{\pm} (\mathrm{pions}), W^{\pm}, Y^{\pm}$ and $Z^{\pm} (\mathrm{kaons})$ are varied in this analysis, see Barr et al.~\cite{PhysRevD.74.094009} for a detailed description.
%The latter scale the kaon and pion production rates for different regions in the phase-space of primary energy and energy transfer $x_{\mathrm{lab}}$. 
With this implementation, uncertainties on the atmospheric neutrino flux predictions induced by hadronic interaction models are covered. As a result, more freedom is given to the fit as can be seen for instance in figure \ref{fig:barrsystematics} where the changes in the zenith and energy distributions introduced by the $W^{-}$-parameter are shown.\\
%To cover uncertainties arising from the primary cosmic-ray flux, a nuisance parameter is implemented in the fit which interpolates continuously between the H4a and GST-4gen models.

\section{Results}
\subsection{Updated Astrophysical Flux Fit Results}
The updated best-fit astrophysical component, assuming a single power-law in energy, is given as:

\begin{equation}
\frac{d\phi_{\nu + \bar\nu}}{dE} = ( 1.44 ^{+0.25}_{-0.24}) \left ( \frac{E}{100\,\mathrm{TeV} }\right )^{-2.28 ^{+0.08}_{-0.09}}\cdot 10^{-18}\,\mathrm{GeV}^{-1}\mathrm{cm}^{-2}\mathrm{s}^{-1}\mathrm{sr}^{-1}.
\end{equation}

The central range of neutrino energies that contribute 90\% to the total observed likelihood ratio between the best-fit and the atmospheric-only hypothesis in the experimental data is $40\,\mathrm{TeV} - 3.5\,\mathrm{PeV}$. The result is consistent with the one previously reported based on eight years of data~\cite{ICRC17NuMu}, but a softening of the best-fit spectral index is observed. This change is mostly caused by the more accurate treatment of the primary cosmic-ray flux in MCEq compared to the approximate re-weighting that was applied in previous iterations of the analysis~\cite{AnnePHD}.

\begin{figure}[htbp]
		\includegraphics[width=0.94\textwidth]{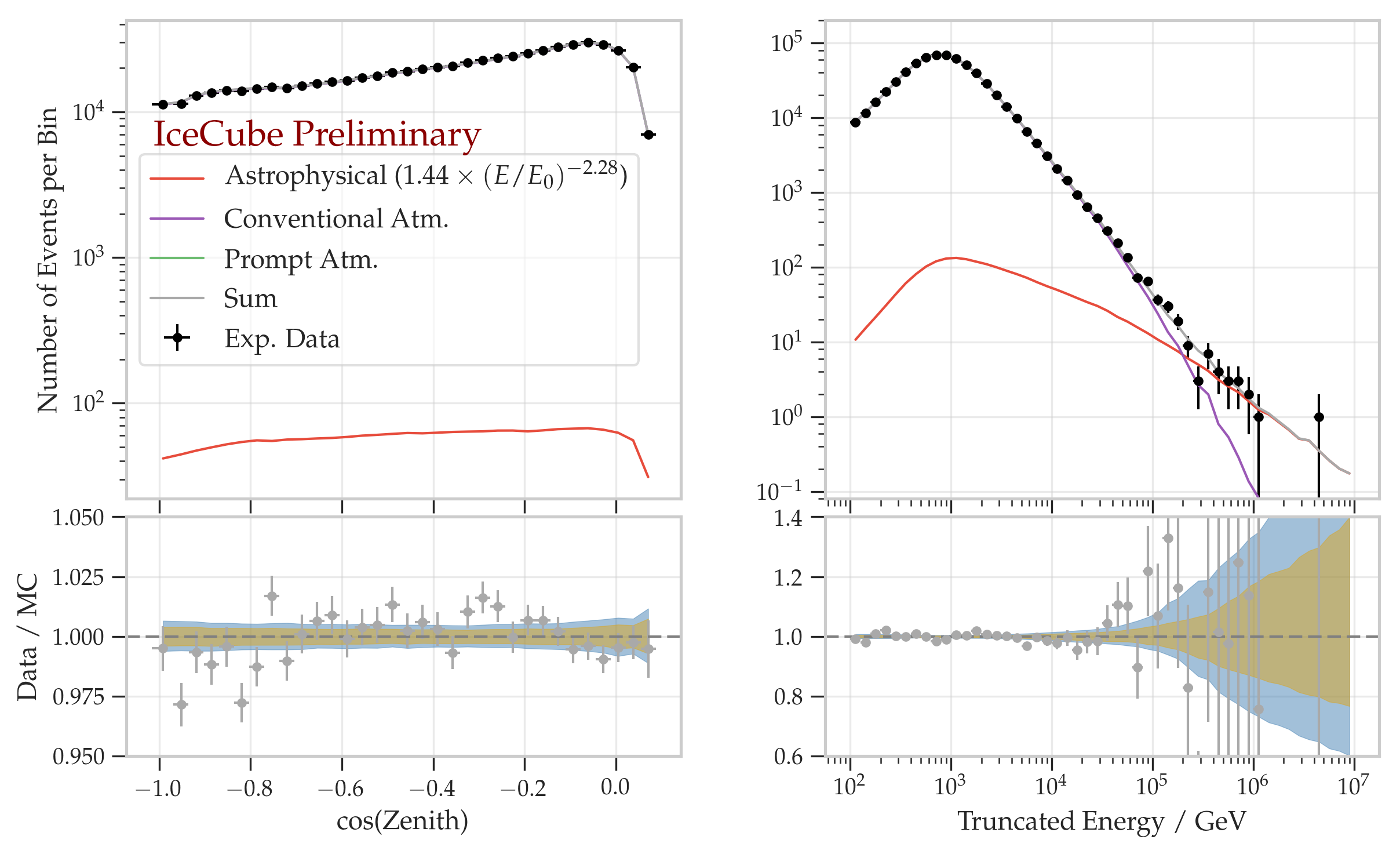}
 \caption{Data from 2010 to 2018 together with the best-fit expectation from Monte Carlo simulations. Left: Distribution of the cosine of the reconstructed zenith. Right: Distribution of the reconstructed muon energy. The brown and blue bands mark the central $68\%$ and $90\%$ spread of the expectation if all fit-parameters are varied within their posterior distribution ranges, taking their correlations into account.}
 \label{fig:data_vs_bestfit}
\end{figure}

Figure \ref{fig:data_vs_bestfit} shows the comparison of the best-fit expectation with the experimental data from 2010 to 2018 (Pass-2) as a function of reconstructed observables. The data from 2009 (IC59) is fitted simultaneously, but not shown here. A small deficit of about $1-2\%$ is observed for straight up-going events ($\cos(\theta)<-0.8$) between the data and the best-fit expectation. This effect is driven by events with energies smaller than $5\,\mathrm{TeV}$ and is currently under investigation. It does however not strongly affect the astrophysical measurement, as confirmed by repeating the fit excluding events with $\cos(\theta)<-0.7$. 

\section{Discussion}
\begin{figure}[htp]
	\centering
	\includegraphics[width=0.78\textwidth]{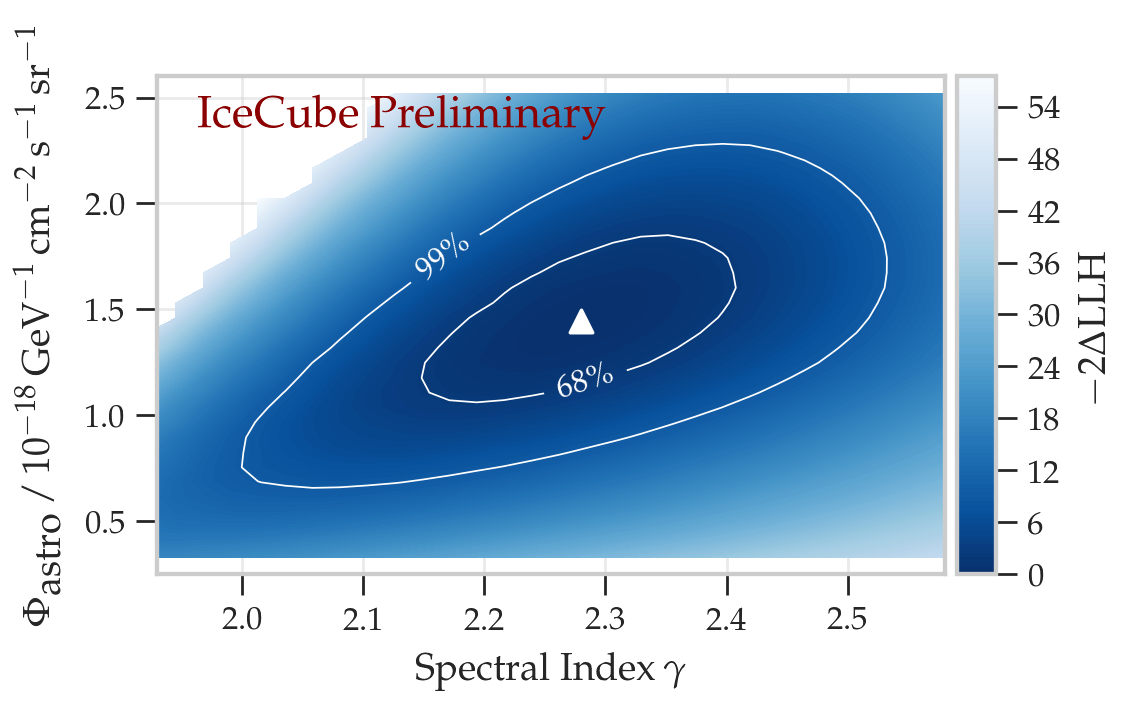}
	\caption{Scan of the profile likelihood for the two signal parameters: astrophysical normalization and spectral index. Note that for each scan point, all other parameters are optimized. \label{fig:scans}}
\end{figure}

The two-dimensional contour of the profile likelihood as a function of the signal parameters is shown in figure \ref{fig:scans}. It shows the expected correlation between the astrophysical flux normalization and its spectral index. All other fit-parameters, including the normalization of the prompt atmospheric flux, show very little correlation with the astrophysical signal parameters.

\subsection{Prompt Atmospheric Neutrinos}
The sub-dominant component of prompt atmospheric neutrinos is currently being investigated in more detail. The best-fit normalization is zero, and the effort to provide an updated flux upper limit is ongoing. On the other hand, even if a non-zero flux of prompt neutrinos is assumed as a benchmark case, the impact on the astrophysical parameters is small and an astrophysical flux would remain necessary to describe the data: Fixing the prompt normalization in the fit to the baseline prediction from MCEq (Sec.~\ref{sec:neutrinofluxes}) hardens the astrophysical spectral index by $\Delta \gamma_{astro} = -0.05$ and the astrophysical normalization decreases to $1.17\cdot 10^{-18}\,\mathrm{GeV}^{-1}\mathrm{cm}^{-2}\mathrm{s}^{-1}\mathrm{sr}^{-1}$. An important point to note is that the extrapolation of the unbroken power-law as model for the astrophysical component to lower energies restricts the potential contribution from a prompt component in the medium energy region ($E_{\nu} \approx 10\,$TeV), more complex models for the astrophysical component are currently being studied. 

\subsection{Primary Cosmic-Ray Flux}
The primary cosmic-ray flux is an important ingredient for the calculation of the atmospheric neutrino fluxes. Its exact shape and composition are uncertain, especially in the transition energy region where both the atmospheric and the astrophysical flux contribute significantly to the total flux. To estimate the impact of this uncertainties on the measured astrophysical flux, various models have been used to calculate the atmospheric neutrino fluxes and the fit has been repeated for each of them, see table \ref{tab:CR_impact} for results. The astrophysical flux normalization and spectral index do not change strongly between the different CR-models, most differences are absorbed by nuisance parameters, especially $\Delta \gamma$ which varies the spectral index of the primary CR spectrum. However, the likelihood difference to the baseline model H4a (last column) shows that overall the data is best described if the GST-4gen model is used. 

\begin{table}[h!]
\centering
{\small
\begin{tabular}{lcccc}
\hline
{CR-Model} & {Change of $\gamma_\mathrm{astro}$} & {Change of $\Phi_\mathrm{astro}$} & {$\Delta \gamma_\mathrm{CR}$} & {$\Delta LLH$}\\
\hline
        Gaisser-Hillas, H4a\cite{Gaisser:2012_H3a_n_H4a} & -- & -- & 0.06 & -- \\
        Gaisser-Stanev-Tilav, GST-4gen \cite{Gaisser:2013bla} & -0.007 & -0.115 & 0.01 & 4.4 \\
        GSF-beta \cite{Dembinski:2017zsh} & -0.007 & -0.234 & 0.03 & 1.8 \\
        Mascaretti et al. (KASCADE w. cutoff) \cite{Mascaretti:2019mnk} & 0.015 & -0.235 & -0.04 & -2.1 \\
        Mascaretti et al. (ARGO-YBJ w. cutoff) \cite{Mascaretti:2019mnk}  & -0.011 & 0.116 & 0.02 & -1.9 \\
        \hline
\end{tabular}\\
}
\caption{Impact of different primary cosmic-ray flux models (CR-models) used for the calculation of atmospheric fluxes on the astrophysical fit-parameters compared to the baseline model H4a. \label{tab:CR_impact}}
\end{table}

\section{Conclusions and Outlook}
\begin{figure}[htbp]
    \centering
    \includegraphics[width=0.94\textwidth]{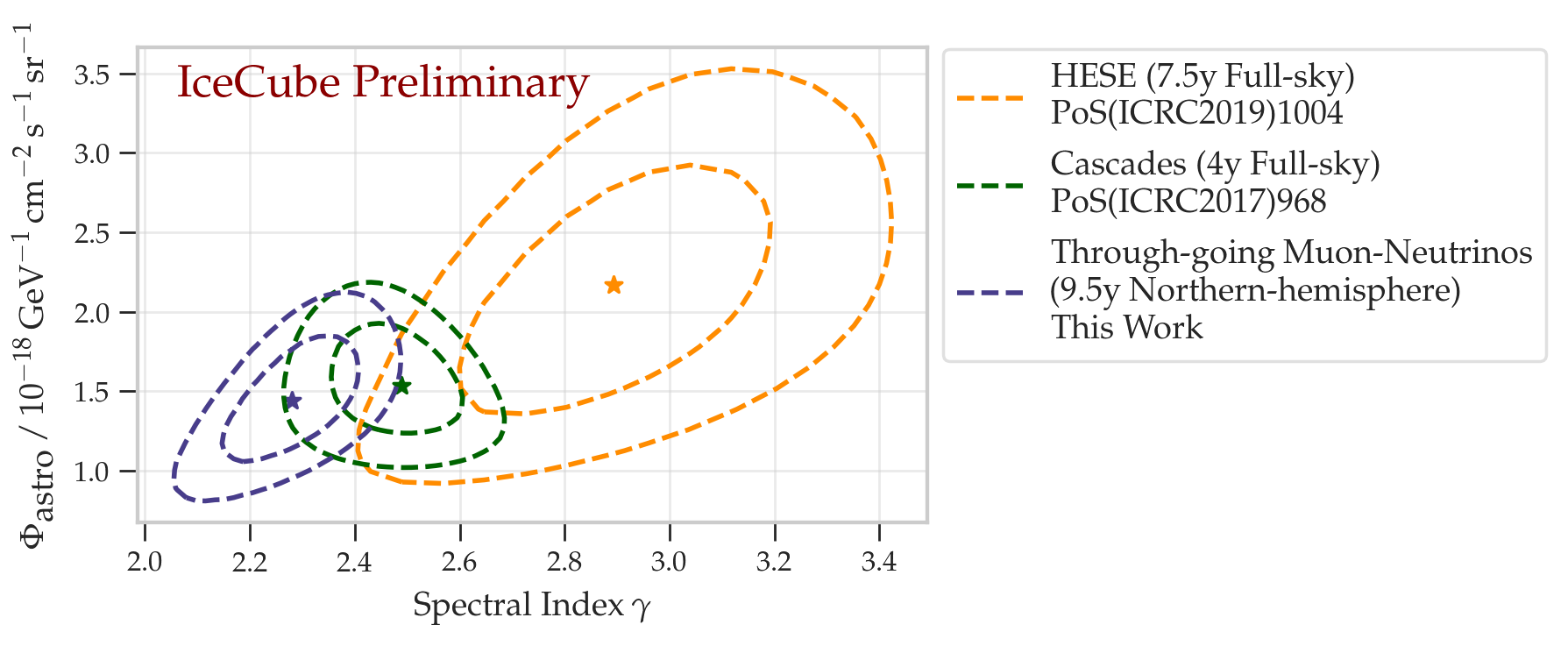}
    \caption{Comparison of the best-fits and profile likelihood contours ($68\%$ and $95\%$ CL) of different IceCube analyses measuring the astrophysical flux, assuming a single power-law energy spectrum. The y-axis shows the per-flavor normalization anchored at $100\,$TeV.}
    \label{fig:ICcomparison}
\end{figure}

We have presented an update of the analysis of muon neutrinos from the Northern Hemisphere, now based on almost 10 years of IceCube operation. The treatment of systematic uncertainties, especially with respect to uncertainties on the atmospheric fluxes, has been improved and most of the experimental data has been re-calibrated and re-processed to obtain a consistent dataset (Pass-2). A softening of the single power-law is observed with respect to the values reported in \cite{ICRC17NuMu}, the updated best-fit parameters are $\Phi_{astro}=1.44 \pm^{0.25}_{0.24}$ and $\gamma_{astro}=2.28 \pm^{0.08}_{0.09}$. However, these are consistent with the previously reported values. Figure \ref{fig:ICcomparison} shows a comparison with other IceCube measurements of the diffuse astrophysical neutrino flux, namely those resulting from high-energy starting events \cite{ICRC19HESE:GlobalFit} and cascade-like events \cite{ICRC17CASCADES}. 

We want to stress that this update focused on the single power-law as baseline model, but more complex models for the astrophysical component are currently being tested.
%and an updated result using all IceCube datasets is in preparation.
\bibliographystyle{ICRC}

\bibliography{references}

\providecommand{\href}[2]{#2}\begingroup\raggedright\begin{thebibliography}{10}

\bibitem{Aartsen:2013jdh}
{\bf IceCube} Collaboration, M.~G. Aartsen et~al., {\em Science} {\bf 342}
  (2013) 1242856.

\bibitem{Aartsen:2015rwa}
{\bf IceCube} Collaboration, M.~G. Aartsen et~al., {\em Phys. Rev. Lett.} {\bf
  115} (2015) 081102.

\bibitem{Aartsen:2016xlq}
{\bf IceCube} Collaboration, M.~G. Aartsen et~al., {\em Astrophys. J.} {\bf
  833} (2016) 3.

\bibitem{Aartsen:2016nxy}
{\bf IceCube} Collaboration, M.~G. Aartsen et~al., {\em JINST} {\bf 12} (2017)
  P03012.

\bibitem{ICRC17NuMu}
{\bf IceCube} Collaboration,  \pos{PoS(ICRC2017)1005} (2017).

\bibitem{GNCalert_EHE20171106}
{\bf IceCube} Collaboration, M.~G. Aartsen et~al., {\em GRB Coordinates
  Network, Circular} (2017). https://gcn.gsfc.nasa.gov/gcn3/22105.gcn3.

\bibitem{Fedynitch:2015zma}
A.~Fedynitch, R.~Engel, T.~K. Gaisser, F.~Riehn, and T.~Stanev, {\em EPJ Web
  Conf.} {\bf 99} (2015) 08001.

\bibitem{MSIS00_2002JGRA}
J.~M. {Picone}, A.~E. {Hedin}, D.~P. {Drob}, and A.~C. {Aikin}, {\em Journal of
  Geophysical Research (Space Physics)} {\bf 107} (Dec., 2002) 1468.

\bibitem{Fedynitch:2018cbl_SIBYLL23c}
A.~Fedynitch, F.~Riehn, R.~Engel, T.~K. Gaisser, and T.~Stanev,
  \href{http://arxiv.org/abs/1806.04140}{{\tt arXiv:1806.04140}}.

\bibitem{Gaisser:2012_H3a_n_H4a}
T.~K.~Gaisser, {\em Astroparticle Physics} {\bf 35} (07, 2012) 801--806.

\bibitem{bibWilks}
S.~S. Wilks, {\em Ann. Math. Statist.} {\bf 9} (1938) 60--62.

\bibitem{PhysRevD.74.094009}
G.~D. Barr, S.~Robbins, T.~K. Gaisser, and T.~Stanev, {\em Phys. Rev. D} {\bf
  74} (Nov, 2006) 094009.

\bibitem{AnnePHD}
A.~Schukraft.
\newblock PhD thesis, RWTH Aachen University, 2013.

\bibitem{Gaisser:2013bla}
T.~K. Gaisser, T.~Stanev, and S.~Tilav, {\em Front. Phys.} {\bf 8} (2013)
  748--758.

\bibitem{Dembinski:2017zsh}
H.~P. Dembinski, R.~Engel, A.~Fedynitch, T.~Gaisser, F.~Riehn, and T.~Stanev,
  \pos{PoS(ICRC2017)533} (2018). [35,533(2017)].

\bibitem{Mascaretti:2019mnk}
C.~Mascaretti, P.~Blasi, and C.~Evoli,
  \href{http://arxiv.org/abs/1906.05197}{{\tt arXiv:1906.05197}}.

\bibitem{ICRC19HESE:GlobalFit}
{\bf IceCube} Collaboration,  \pos{PoS(ICRC2019)1004}  (these proceedings).

\bibitem{ICRC17CASCADES}
{\bf IceCube} Collaboration,  \pos{PoS(ICRC2017)968} (2017).

\end{thebibliography}\endgroup

\end{document}